\documentclass[
 reprint,
 amsmath,
 amssymb,
 superscriptaddress,
 aps,
]{revtex4-1}

\usepackage{graphicx}
\usepackage{bm}
\usepackage{dcolumn}
\usepackage{gensymb}
\usepackage[utf8]{inputenc}
\usepackage{float}

\usepackage{array,mathtools,amssymb,booktabs}
\newcolumntype{C}{>{$}c<{$}}
\AtBeginDocument{
\heavyrulewidth=.08em
\lightrulewidth=.05em
\cmidrulewidth=.03em
\belowrulesep=.65ex
\belowbottomsep=0pt
\aboverulesep=.4ex
\abovetopsep=0pt
\cmidrulesep=\doublerulesep
\cmidrulekern=.5em
\defaultaddspace=.5em
}

\usepackage{hyperref}
\begin{document}

\title{Predicting the energies of Cf$^{17+}$ for an optical clock}

\author{S. G. Porsev}
\affiliation{Department of Physics and Astronomy, University of Delaware, Newark, Delaware 19716, USA}
\author{M. S. Safronova}
\affiliation{Department of Physics and Astronomy, University of Delaware, Newark, Delaware 19716, USA}

\begin{abstract}
Highly charged ions (HCIs) combine compact electronic structure with strong relativistic effects, offering both robustness against external perturbations and enhanced sensitivity to variations of the fine-structure constant.  Recent advances in sympathetic cooling and trapping enable precision measurements of highly charged ions; however, fully exploiting their potential requires accurate theoretical predictions. In particular, reliable calculations of clock wavelengths are essential for experimentally locating HCI clock transitions.
Here, we treat Cf$^{17+}$ as a univalent ion and perform calculations within the relativistic coupled-cluster framework, iteratively including nonlinear single–double contributions and valence and core triple excitations. We also assess quantum-electrodynamic corrections and basis-set and partial-wave truncation effects. Our results establish the impact of different correlation contributions on the low-lying energy spectrum and provide a quantitatively reliable prediction of the $5f_{5/2}\to 6p_{1/2}$ clock transition, highlighting the critical role of core–valence correlations and iterative triples for precision spectroscopy and optical clock development.
\end{abstract}

\date{\today}

\maketitle

\section{Introduction}
Highly charged ions (HCIs) provide an exceptional platform for high-precision spectroscopy and next-generation optical clocks. Their compact electronic structure suppresses sensitivity to external perturbations, while strong relativistic effects and large ionization energies enhance sensitivity to variations of the fine-structure constant $\alpha$~\cite{BerDzuFla10}. These features make HCIs uniquely suited for precision tests of fundamental physics, including searches for hypothetical oscillatory variation of $\alpha$, such as those induced by topological defects or cosmological fields, including dark matter \cite{DerPos14,StaFla15,StaFla16}.

Experimental progress toward highly charged ion (HCI) optical clocks has accelerated with advances in sympathetic cooling, quantum-logic spectroscopy, and rapid laser frequency scanning techniques \cite{SchVerSch15,KozSafCre18,MicLeoKin20,CheSpiWil24}. An optical clock based on Ar$^{13+}$ has been demonstrated with a total systematic uncertainty at the $2\times10^{-17}$ level~\cite{King2022Nature}, establishing HCIs as competitive frequency standards. More recently, the ultra-narrow Ni$^{12+}$ 
$^3\!P_{2} \to\, ^3\!P_{0}$ electric-quadrupole clock transition was successfully located using high-accuracy \emph{ab initio} predictions \cite{Cheung2025Ni12}, underscoring the critical role of reliable theory in identifying previously unobserved HCI clock transitions.

Fully exploiting HCIs therefore, requires a quantitatively reliable treatment of electron-correlation effects, including the systematic inclusion of core–valence correlations and higher-order triple-excitation contributions.
Comparisons with high-precision measurements provide critical benchmarks for assessing theoretical uncertainties and are especially important for predicting the properties of very heavy elements before challenging spectroscopic studies~\cite{PorSafSaf18}.

In earlier work \cite{PorSafSaf20}, we assessed the feasibility of an optical clock based on the $6s^2\, 5f_{5/2} \to 6s^2\, 6p_{1/2}$ transition in Cf$^{17+}$, treating this ion as a trivalent system and calculating the relevant energies, hyperfine-structure constants, electric-quadrupole shifts, and polarizabilities.
That study highlighted the pivotal role of core–valence correlations and iterative triple excitations in determining the clock transition. Several experimental programs are pursuing optical-clock transitions in Cf ions, making it essential to predict the Cf$^{17+}$ clock-transition energy with the highest possible accuracy to enable reliable experimental searches. This motivates a renewed theoretical investigation employing an alternative computational approach.

Here, we treat Cf$^{17+}$ as a univalent ion and perform calculations within the relativistic coupled-cluster framework, which efficiently captures core–core and core–valence correlations. This approach simplifies the treatment of triple excitations while maintaining high accuracy. Starting from the linearized coupled-cluster single–double (LCCSD) method, we systematically include nonlinear single–double contributions and valence and core triple excitations by iteratively solving the coupled-cluster single–double–triple (CCSDT) equations. This allows an essentially complete treatment of linear triple excitations with minimal restrictions. We also assess quantum-electrodynamic corrections and investigate the effects of truncating the partial-wave expansion and limiting the basis set.

The primary goal of this work is to quantify the impact of different correlation contributions on the low-lying energy levels and to provide a reliable prediction of the Cf$^{17+}$ clock transition energy, with particular emphasis on core–valence and iterative triple excitations.
\section{Basis set and method of calculation}
\label{method}
We evaluated the energies of the lowest-lying states using a high-precision relativistic coupled-cluster method developed in Ref.~\cite{PorDer06}.
Triple excitations were included by solving the coupled-cluster equations iteratively. Quadratic nonlinear single and double terms ($S^2$, $SD$, and $D^2$) were also included, while cubic and higher-order nonlinear terms, as well as nonlinear triple terms, were omitted.

Treating Cf$^{17+}$ as a univalent ion, we constructed the basis set in the $V^{N-1}$ approximation, where $N$ is the total number of electrons.
The initial self-consistent-field procedure, including the Breit interaction, was carried out for the core electrons [$1s^2,\ldots,6s^2$]. The $5f$, $6$--$7p$, $6d$, $7s$, and $5g$ orbitals were then constructed in the frozen-core potential.

The virtual $6h$ and $7i$ orbitals were formed using 40 B-spline basis orbitals of order 7 defined on a nonlinear grid with 500 points.
The remaining virtual orbitals were generated using the recurrent procedure described in Refs.~\cite{KozPorFla96,KozPorSaf15}. In this approach, the large component of the radial Dirac bispinor, $f_{n'l'j'}$, is obtained from a previously constructed function $f_{nlj}$ by multiplying it by $r^{l'-l}\sin(kr)$, where $l'$ and $l$ are the orbital angular momenta of the new and old orbitals ($l' \ge l$), and the coefficient $k$ is determined by the properties of the radial grid. The small component, $g_{n'l'j'}$, is obtained from the kinetic-balance condition. The newly constructed functions were orthonormalized with respect to orbitals of the same symmetry.
The basis set included partial waves with orbital angular momentum up to $l=6$.

The coupled-cluster equations were solved in a basis consisting of single-particle states.
In the equations for singles, doubles, and valence triples, summations over excited states were carried out using 35 basis orbitals with $l \le 6$.
Quantum electrodynamic (QED) corrections were included following Refs.~\cite{ShaTupYer13,TupKozSaf16}.

\section{Results and discussion}
Numerical results for the excitation energies are presented in Table~\ref{Tab:E}.
\begin{table}[ht]
\caption{Contributions to excitation energies of the $6p_{1/2}$, $5f_{7/2}$, and $6p_{3/2}$ states (in cm$^{-1}$) in
different approximations, discussed in the text, are presented. The theoretical CCSDT values are given in the row $E_{\rm CCSDT}$.
The QED and extrapolation corrections are given in the rows
$\Delta E_{\rm QED}$ and $\Delta E_{\rm extrap}$.
The results of other works are listed in the bottom panel for comparison.}
\label{Tab:E}
\begin{ruledtabular}
\begin{tabular}{lrrr}
\smallskip
                                                 & $6p_{1/2}$ & $5f_{7/2}$ &    $6p_{3/2}$   \\
\hline \\[-0.7pc]
$E_{\rm BDHF}$                                   &  12726     &   19835    &     235781      \\[0.3pc]
$\Delta E_{\rm SD}$                              &   7948     &    1056    &       7055      \\[0.1pc]
$E_{\rm LCCSD}$                                  &  20674     &   20891    &     242836      \\[0.2pc]
$E_{\rm CI+LCCSD}$~\cite{PorSafSaf20}$^{\rm a}$  &  20716     &   21014    &     243287      \\[0.4pc]
$\Delta E_{\rm vNL}$                             &     74     &       2    &         81      \\[0.1pc]
$\Delta E_{\rm cNL}$                             & $-$145     &    $-$4    &         14      \\[0.2pc]
$\Delta E_{\rm vT}$                              & $-$688     &   $-$68    &     $-$249      \\[0.1pc]
$\Delta E_{\rm cT}$                              &  $-$56     &       3    &     $-$281      \\[0.1pc]
$E_{\rm CCSDT}$                                  &  19856     &   20825    &     242402      \\[0.4pc]
\multicolumn{3}{l}{\it Complementary corrections:} \\
{$\Delta E_{\rm QED}$}                           &    631     &      12    &        662      \\[0.1pc]
$\Delta E_{\rm extrap}$                          &    970     &      33    &       1060      \\[0.2pc]
\hline \\[-0.6pc]
$E_{\rm total}$                                  &  21456     &   20870    &     244125      \\[0.3pc]
Other works:                                     &&& \\
Porsev {\it et al.}~\cite{PorSafSaf20}           &  20611     &   20895    &     243081      \\[0.1pc]
Berengut {\it et al.}~\cite{BerDzuFla12}         &  18686     &   21848    &     242811
\end{tabular}
\end{ruledtabular}
\begin{flushleft}
$^{\rm a}$These values are obtained as the sum of the CI+MBPT contributions  and the HO corrections from Ref.~\cite{PorSafSaf20}.
\end{flushleft}
\end{table}
The lowest-order Dirac–Hartree–Fock contribution to the excitation energies (measured from the ground $5f$ state) is labeled ``BDHF,'' reflecting the inclusion of the Breit interaction.
The correction $\Delta E_{\rm SD}$ is defined as the difference between the LCCSD and BDHF values shown in the third and first rows of Table~\ref{Tab:E}.

The fourth row gives the CI+LCCSD values from Ref.~\cite{PorSafSaf20}, obtained by adding the CI+MBPT contributions and
the higher-order (HO) corrections listed in Table~I of that work. The close agreement with the present LCCSD results, especially for the $6p_{1/2}$ excitation energy, shows that configuration interaction effects are small for these states and validates the univalent treatment of Cf$^{17+}$.

We then performed calculations in the CCSDT approximation, including valence and core nonlinear terms (vNL and cNL), as well as valence and core triple excitations (vT and cT), yielding the corrections $\Delta E_{\rm v(c)NL}$ and $\Delta E_{\rm v(c)T}$, respectively.
The nonlinear and triple contributions were evaluated separately.
We verified that the interference between nonlinear and triple terms is small and can be neglected.

As expected, the correlation corrections ($\Delta E_{\rm SD}$, $\Delta E_{\rm v(c)NL}$, and $\Delta E_{\rm v(c)T}$) are dominated by the SD contribution. The CCSDT energies were obtained as
\begin{equation}
E_{\rm CCSDT} = E_{\rm LCCSD} + \Delta E_{\rm NL} + \Delta E_{\rm T},
\end{equation}
where $\Delta E_{\rm NL} \equiv \Delta E_{\rm vNL} + \Delta E_{\rm cNL}$ and
$\Delta E_{\rm T} \equiv \Delta E_{\rm vT} + \Delta E_{\rm cT}$.
We note that both the nonlinear and triple corrections are substantial. For the $6p_{1/2}$ and $5f_{7/2}$ states, $\Delta E_{\rm NL}$ and $\Delta E_{\rm vT}$ have the same sign and therefore add rather than cancel.

Because Cf$^{17+}$ is a highly charged ion, the energies for the valence $5f_{5/2}$ and $6p_{1/2}$ states are very large in absolute value and close in magnitude:
$E(5f_{5/2}) \approx -2.623\times10^6~{\rm cm}^{-1}$ and
$E(6p_{1/2}) \approx -2.602\times10^6~{\rm cm}^{-1}$.
As a result, the transition energy lies in the optical range.
The nonlinear and triple corrections to these energies are small: the combined valence and core triple contributions do not exceed 0.06\% for any of the considered states, while the combined nonlinear contributions are below 0.01\%.
Nevertheless, their effect on transition energies is non-negligible and must be taken into account.

Additional corrections due to basis-set extrapolation ($\Delta E_{\rm extrap}$) and QED radiative effects ($\Delta E_{\rm QED}$) were also evaluated.
To determine the extrapolation corrections, we constructed an extended basis including eight partial waves ($l_{\rm max}=7$) and orbitals with principal quantum number up to $n=35$, and performed calculations within the LCCSD framework.
Following the empirical rule established for Ag-like ions in Ref.~\cite{SafDzuFla14PRA1} and used in previous studies~\cite{PorCheSaf24}, the contribution of higher partial waves ($l>6$) was estimated as twice the difference between results obtained with $l_{\rm max}=7$ and $l_{\rm max}=6$.

The QED and extrapolation corrections for the $6p_{1/2}$ and $6p_{3/2}$ states are large and comparable in magnitude to $\Delta E_{\rm T}$, making their inclusion essential.
Since $\Delta E_{\rm extrap}$ and $\Delta E_{\rm T}$ have opposite signs for all considered states, they partially cancel.

The final theoretical values, denoted by $E_{\rm total}$, were obtained by adding $\Delta E_{\rm QED}$ and $\Delta E_{\rm extrap}$ to the CCSDT energies.
These results are in reasonable agreement with Ref.~\cite{PorSafSaf20}, but differ noticeably from the values reported by Berengut \textit{et al.}~\cite{BerDzuFla12}.
The primary source of this discrepancy is the omission of the Breit interaction in Ref.~\cite{BerDzuFla12}, which is particularly important in this system.
For example, it shifts the excitation energy of the $6p_{1/2}$ state by more than $4000~{\rm cm}^{-1}$.

Given the high computational cost of iterative solutions for valence (and especially core) triples, it is useful to assess the relative importance of different contributions.
To this end, we solved the triple-excitation equations using several levels of truncation:

(i) for valence triples, excitations were allowed from the $[3s$–$6s]$ core shells with $l^{\rm tr}_{\rm max}=5$ and $n^{\rm tr}_{\rm max}=20$; for core triples, excitations were allowed from the $[4s$–$6s]$ core shells with $l^{\rm tr}_{\rm max}=4$ and $n^{\rm tr}_{\rm max}=20$;

(ii) the same as (i), but with $n^{\rm tr}_{\rm max}=30$ and core excitations allowed from $[3s$–$6s]$;

(iii) full inclusion of valence triples, with excitations from all core shells, $l^{\rm tr}_{\rm max}=6$, and $n^{\rm tr}_{\rm max}=35$.

The remaining contribution from core excitations to $n>30$ and $l=5$–6 is small and can be neglected within our target accuracy.

The resulting valence and core triple contributions to the excitation energies are listed in Table~\ref{E:trip}.
\begin{table}[ht]
\caption{The valence (vT) and core (cT) triple-excitation contributions to the excitation energies (in cm$^{-1}$) are listed. The second column specifies $n_{\rm max}^{\rm tr}$, $l_{\rm max}^{\rm tr}$, and the core shells from which excitations were allowed for cases (i)–(iii), described in detail in the main text.}
\label{E:trip}
\begin{ruledtabular}
\begin{tabular}{lcrrr}
\smallskip
                       & $n_{\rm max}^{\rm tr}-l_{\rm max}^{\rm tr}- [a-b]$
                                          & $6p_{1/2}$ &    $5f_{7/2}$    & $6p_{3/2}$  \\
\hline \\[-0.7pc]
$\Delta E_{\rm vT}$    & (i  = 20-5-$[4s-6s]$) &    -277    &        -40       &     269     \\[0.1pc]
                       & (ii = 30-5-$[3s-6s]$) &    -694    &        -61       &    -248     \\[0.1pc]
                       & (iii= 35-6-$[1s-6s]$) &    -688    &        -68       &    -249     \\[0.4pc]
$\Delta E_{\rm cT}$    & (i = 20-4-$[4s-6s]$)  &     -72    &          3       &    -280     \\[0.1pc]
                       & (ii= 30-4-$[3s-6s]$)  &     -56    &          3       &    -281
\end{tabular}
\end{ruledtabular}
\end{table}
To estimate the uncertainty of the $6p_{1/2}$ excitation energy, we note that the linear singles and doubles contribution shifts the BDHF energy by $7948~{\rm cm}^{-1}$.
The combined nonlinear and triple correction amounts to $-818~{\rm cm}^{-1}$, approximately 10\% of the SD contribution.
Assuming conservatively that omitted higher-order terms (such as cubic nonlinear terms and nonlinear triples) contribute at the level of 30\% of this correction, we estimate this contribution as $d_{\rm NLT} \approx 240$–$250~{\rm cm}^{-1}$.
The uncertainty of the QED correction is estimated as 10\%, yielding $d(\Delta E_{\rm QED}) \approx 60~{\rm cm}^{-1}$.

An additional uncertainty arises from the extrapolation correction.
In Ref.~\cite{PorSafSaf20}, this correction for the $6p_{1/2}$ state, found to be $1021~{\rm cm}^{-1}$, can be compared with the present value $970~{\rm cm}^{-1}$.
We therefore estimate the uncertainty $d(\Delta E_{\rm extrap}) \approx 50~{\rm cm}^{-1}$ as the difference between these two values.

The total uncertainty of the $6p_{1/2}$ excitation energy is then
\begin{equation}
dE(6p_{1/2}) =
\sqrt{
d_{\rm NLT}^2 +
d(\Delta E_{\rm QED})^2 +
d(\Delta E_{\rm extrap})^2
},
\end{equation}
yielding $dE(6p_{1/2}) \approx 250~{\rm cm}^{-1}$.
\section{Conclusion and final remarks}
We have carried out relativistic coupled-cluster calculations of the excitation energies of the low-lying $6p_{1/2}$, $5f_{7/2}$, and $6p_{3/2}$ states of Cf$^{17+}$ using the CCSDT method. The present work includes an iterative solution of the coupled-cluster equations for both valence and core triple excitations, together with the quadratic nonlinear single–double terms. This provides a near-complete treatment of linear triple excitations for a heavy univalent ion.

A detailed analysis of individual correlation contributions demonstrates that, while the dominant corrections arise from linearized single and double excitations, both nonlinear and triple-excitation terms give substantial contributions that do not cancel in this system.
We also show that basis-set extrapolation and QED radiative corrections are comparable in magnitude to the triple-excitation contributions for the $6p_{1/2}$ and $6p_{3/2}$ states and must be included for reliable results.

The resulting excitation energies are in reasonable agreement with previous CI+all-order calculations but differ significantly from earlier results that neglected the Breit interaction. An uncertainty estimate for the $6p_{1/2}$ excitation energy yields an uncertainty of approximately 250~cm$^{-1}$.

The present study establishes the quantitative role of triple excitations in high-precision calculations for heavy univalent systems. The methodology and conclusions of this work can be directly extended to other highly charged ions and heavy atoms, where experimental data are scarce and reliable theoretical predictions are required.

\section*{Acknowledgments}
This work has been supported in part by NSF Award 2309254, US Office of Naval Research Grant N000142512105, and the European Research Council (ERC) under
the European Union's Horizon 2020 research and innovation program (Grant Agreement No. 856415).


%

\end{document}